# Anisotropic magnetotransport of electron gases at SrTiO$_3$ (111) and (110) surfaces with high mobility

Ludi Miao,[1] Renzhong Du,[1] Yuewei Yin[1] and Qi Li[1,a]

[1]*Department of Physics, The Pennsylvania State University, University Park, Pennsylvania 16802, USA*

## Abstract

Electron gases at the surfaces of (001), (110), and (111) oriented SrTiO$_3$ (STO) have been created using Ar$^+$-irradiation with fully metallic behavior and low-temperature-mobility as large as 5500 cm$^2$V$^{-1}$s$^{-1}$, 1300 cm$^2$V$^{-1}$s$^{-1}$ and 8600 cm$^2$V$^{-1}$s$^{-1}$ for (001)-, (110)-, and (111)-surfaces, respectively. The in-plane anisotropic magnetoresistance (AMR) have been studied for the samples with the current along different crystal axis directions to subtract the Lorentz Force effect. The AMR shows features which coincide with the fixed orientations to the crystalline axes, with 4-fold, 2-fold and nearly-6-fold symmetries for (001)-, (110) and (111)-surfaces, respectively, independent of the current directions. These features are possibly caused by the polarization of spin orbit texture of the 2D Fermi surfaces. In addition, a 6-fold to 2-fold symmetry breaking for (111)-surfaces is observed. Our results demonstrate the effect of symmetry of two-dimensional electronic structure on the transport behaviors for the electron gases at STO surfaces.

---

Author to whom correspondence should be addressed. Electronic address: [a]qil1@psu.edu

# I. Introduction

The discovery of two-dimensional electron gases (2DEGs) at oxide surfaces [1] and interfaces [2-3] has opened up broad interests in both condensed matter physics and electronic device applications due to their fascinating exotic properties such as quantum Hall effect,[4-5] two-dimensional (2D) superconductivity,[6-9] ferromagnetism [10-13], and gate-tunable ground states.[7-8] Exciting examples are 2DEGs at the surfaces [1,14] and interfaces [2] of $SrTiO_3$ (STO), a transition metal oxide (TMO) whose bulk is an insulator with a large band gap of ~3.5 eV. These include the heterostuctures of STO and another TMO insulator such as $LaAlO_3$ (LAO) [2] and $LaTiO_3$,[15] non-perovskite or amorphous oxide capped STO,[16,17] delta-doped STO,[18] $Ar^+$ ion [1] or synchrotron irradiated [14] STO surfaces, STO single crystals with electrolyte gating,[7] and vacuum-cleaved STO.[19] Although these 2DEG systems involve various possible mechanisms such as polar catastrophe induced charge transfer effect,[20] oxygen deficiencies,[1,14] and electric field effect,[7,8] their similar transport properties point to a universal underlying physics: the electrons that partially fill into the empty Ti $3d^0$ band near the interfaces/surfaces are responsible for the formation of the 2DEGs. To understand the electronic structures of these STO based 2DEGs, measurements including angle-resolved photoemission spectroscopy (ARPES) measurements [14,19,21] and first principle calculations,[21-23] have been made. It is established that the electronic structure consists of multiple subbands of light and heavy electrons,[14,21] with Rashba spin splitting, a consequence of spin-orbit coupling at spatial asymmetric surfaces and interfaces.[21,24]

Motivated by recent theoretical works which predict intriguing phenomena in (111)-oriented perovskites,[25-28] where bilayers of transition metal ions form honeycomb lattices that resemble the structures of graphene and topological insulator $Bi_2Se_3$, much attention has been attracted recently to 2DEGs at surfaces/interfaces of STO and $KTaO_3$ (KTO) with (111) orientation.[29-31] ARPES measurements have revealed that the 2D Fermi surface exhibits 6-fold symmetry, a topology novel to previously known 4-fold symmetry in (001)-oriented 2DEGs, along with distinct orbital ordering of $t_{2g}$ manifold as well as strongly anisotropic effective masses.[29-30] STO (110) based 2DEGs were also measured by ARPES, displaying a Fermi surface with 2-fold symmetry and high electronic anisotropy.[30,32] However, how the anisotropic electronic structure of STO (111) and (110) based 2DEGs affect the anisotropic magnetotransport is yet to be investigated.

We have created electron gases with fully metallic behavior and high mobility at surfaces of STO with (110) and (111) orientations using $Ar^+$-irradiation and systematically measured their magnetotransport properties for the first time. STO with (001) orientation is also studied for comparison. Their in-plane anisotropic magnetoresistance (AMR) exhibits a combination of a component that conveys their 4-fold, 2-fold and 6-fold crystalline symmetry and a Lorentz force effect (LFE) induced 2-fold component. In addition, a 6-fold-to-2-fold symmetry breaking in AMR is observed for STO (111). We will discuss the origin of the AMR component that is associated with the crystalline symmetries. Our results reveal intrinsic correlations between the anisotropic magnetotransport and the electronic structures in 2DEGs at STO surfaces with different symmetries.

**II. Experiment**

Commercial 0.5 mm thick single crystalline undoped STO with (001), (110), and (111) orientations are used in this study. The surfaces of these crystals are patterned into Hall bar structures with channel size 40×20 μm$^2$ using photolithography. They are subsequently irradiated by Ar$^+$ ions at a *dc* acceleration voltage of 500 V on a water-cooled holder with an Ar partial pressure $p_{Ar}$ = 0.35 mTorr at room temperature for 3 minutes. The incident angles of Ar$^+$ ions are 45º tilted from normal directions of surfaces. Ti/Au contacts were then deposited by sputtering and wire-bonded with Au wires with Ohmic contacts. An optical microphotograph of the device is shown in Fig. 1(a), with the highlighted region A representing the Ar$^+$-irradiated area. The resistance and the Hall effect were measured using the standard 5-probe Hall bar method in a physical property measurement system (PPMS, Quantum Design).

**III. Results and Discussions**

Figure 1(b), (c) and (d) summarize the sheet resistance $R_{2D}$, the 2D carrier density $n_{2D}$, and the Hall mobility $\mu_H$ as a function of temperature, respectively, for (001)-, (110)-, and (111)-oriented surfaces. All samples exhibit a fully metallic behavior, with the resistance ratio $R_{300K}/R_{4.2K}$ between room temperature-to-low temperature as high as 300-1300 and residual sheet resistance $R_{2D}$ ($T$ = 4.2K) as low as 7.5-50 Ω. Indeed, $\mu_H$ at 4.2 K achieves high values of 5500 cm$^2$V$^{-1}$s$^{-1}$, 1300 cm$^2$V$^{-1}$s$^{-1}$ and 8600 cm$^2$V$^{-1}$s$^{-1}$ for (001)-, (110)-, and (111)-surfaces, respectively. Compared to the reported values for epitaxial LAO/STO heterostructures and irradiated STO surfaces [1,33-36] the $\mu_H$ values of our samples are among the highest, which is important not only for observing fascinating physical properties, but also for electronics applications. The $n_{2D}$ values in the samples are around $10^{13}$-$10^{14}$ cm$^{-2}$ for all surfaces, the same orders of magnitude as that of LAO/STO (001) heterostructures with 2D behaviors grown under a high oxygen partial pressure, and are 2-3 orders of magnitude lower than that of LAO/STO heterostructures and irradiated STO surfaces with three-dimensional (3D) behaviors.[2,34] These STO surfaces show similar transport behavior to that of 2D LAO/STO (001) heterostructures and thus imply both systems possess similar electronic structures. In 3D LAO/STO heterostructures, one consequence of the large $n_{2D}$ is that the screening effect makes the surface effect such as Rashba spin-orbit coupling not observable.[37] Indeed, the 4-fold in-plane AMR, which is associated with Rashba effect as well as cubic crystalline symmetry, was only observed in the 2D LAO/STO (001), but not in 3D LAO/STO (001).[37] Therefore, we should expect to observe Rashba effect and crystalline symmetry associated AMR for these irradiated STO surfaces.

We have measured in-plane AMR and plotted relative $\Delta$AMR defined by $\Delta AMR(H,\phi)$ = $AMR(H,\phi)$-$<AMR(H,\phi)>$, as functions of azimuthal angle $\phi$ for (001)-, (110)-, (111)-surfaces with various field $H$ at 4.2 K, as shown in Fig. 2(a), (c) and (e). $\phi$ is defined as the relative angle between the current and the magnetic field, $AMR(H,\phi)$ is defined as $[R_{2D}(H,\phi)$-$R_{2D}(0)]/R_{2D}(0)$, and $<AMR(H,\phi)>$ is the average value for AMR over all $\phi$. The insets show the relative directions of Hall bar patterns to the crystalline axes: the currents are along [100], [-110], and [11-2] for (001)-, (110)-, and (111)-surfaces, respectively. The data have been symmetrized using $[\Delta AMR(H,\theta) + \Delta AMR(-H,\theta)]/2 \rightarrow \Delta AMR(H,\theta)$ to remove any effects from contacts or field misalignment.[35,38] The behaviors of relative AMR for these surfaces all have a major contribution from LFE, which is generally 2-fold. However, the detailed features for these relative AMRs are distinct for different surface orientations.

Qualitatively, for (001)-surface, $\Delta AMR$ deviates from LFE-induced 2-fold behavior in lower fields, as four shoulder peaks appear around [110]-equivalent directions, as marked out by arrows in Fig. 2(a). This feature is in line with the 4-fold symmetry for a (001)-surface of the cubic system, similar to the 4-fold AMR observed in 2D LAO/STO (001) heterostructures [37] and the irradiated STO (001) surfaces.[36] For (110)-surface, although $\Delta AMR$ remains 2-fold for all values of $H$, a drifting of maxima directions is observed, as marked out by arrows in Fig. 2(c). This indicates a competition between LFE and another intrinsic mechanism which also gives a 2-fold AMR, consistent with the 2-fold crystalline symmetry of STO (110)-surface. For (111)-surface, six peaks/kinks appear at [11-2], [-211], [1-21] directions as marked out by arrows in Fig. 2(e), matching with the 6-fold crystalline symmetry. In addition, we note that the two peaks at $60^o$ and $240^o$ are stronger than other four peaks/kinks.

Since the LFE effect is related to the angle between the current and the magnetic field, by changing the current directions, we can extract the AMR effect purely from crystalline anisotropy. In Fig. 2(b), (d) and (f) we show $\Delta AMR$ with current directions intentionally tilted by $45^o$, $90^o$ and $30^o$ relatively to the crystalline axes, for (001)-, (110), and (111)-surfaces respectively. As $\phi$ is the relative angle between the current and the magnetic field, these $\Delta AMR(\phi)$ should be regarded as the data measured with fixed current directions but tilted crystalline axes. The detailed features of $\Delta AMR$ show drastic differences compared to the ΔAMR curves discussed above. For (001)-surface, previously observed four shoulder peaks disappear. Instead, two additional shoulder peaks at ~30° and ~210° appear, indicating that the low field 4-fold AMR component for (001)-surface in Fig. 2(a) is associated with crystalline axes. For (110)-surface, the 2-fold behavior is seen for ΔAMR, with the maxima aligned at 90° and 270° for all fields. This is in contrast to the maxima drifting seen in Fig. 2(c), indicating that the intrinsic 2-fold AMR has a direction bound to the crystalline axes. For (111)-surface, six peaks/kinks shift by 30° and follow [11-2], [-211], [1-21] axes, indicating that the 6-fold AMR component is indeed bound to the crystalline axes. Again, we note that two peaks at 90° and 270° are stronger than the other four peaks/kinks. Such a 6-fold to 2-fold symmetry breaking is clearly not a consequence of LFE, as the strong peaks have a relative fixed direction to the crystalline axes.

To further separate intrinsic AMR from LFE-induced AMR quantitatively, Fourier analysis was performed. Furthermore, it can also provide additional quantitative information that could potentially reveal physical properties such as magnetic and orbital information in TMOs.[39,40] Complex $m$-fold Fourier amplitude $A_m$ defined as $AMR(\phi) = \sum A_m \exp(-im\phi)$ as well as $m$-fold squared weight $W_m$ defined by $W_m = |A_m|^2 / \sum |A_k|^2$ are obtained. For (001)- and (111)-surfaces, it seems that intrinsic AMR could be immediately obtained by subtracting the Fourier 2-fold term from the AMR raw data. However, as the mixture of intrinsic AMR and LFE-induced AMR is not a pure superposition, other Fourier terms that are generated by such mixture should also be subtracted. Figure 3 shows the absolute values for Fourier complex amplitudes $|A_m|$ as functions of $H$ at 4.2 K extracted from $\Delta AMR$ measured on samples with various orientations. The insets show relative directions of Hall bar patterns to the crystalline axes. For (001)-surfaces, as shown in Fig. 3(a) and (b), $|A_4|$ and $|A_2|$ both increase with field, corresponding to intrinsic AMR and LFE-induced AMR, respectively. Similarly for (111)-surfaces as shown in Fig. 3(e) and (f), $|A_6|$ should be assigned to intrinsic AMR whereas $|A_2|$ should be assigned to LFE-induced AMR. Interestingly, $|A_6|$ in (001)-surfaces and $|A_4|$ in (111)-surfaces also show a finite value, possibly generated from the mixture as mentioned above. For (110)-surfaces as shown in Fig. 3(c) and (d), $|A_2|$ is dominating, as both intrinsic AMR and LFE-induced AMR are 2-fold.

To further identify the Fourier terms that solely associated with intrinsic AMR and Fourier terms that involved LFE-induced AMR, we looked into the phase of the complex amplitudes $A_m$. Clearly, if one compares AMR with current patterned along different crystalline axes, $A_m$ that solely associated with intrinsic AMR should change phase $\psi_m$ as currents/crystalline axes tilt. Also, the phase change value $\Delta\psi_m$ should be $m$ times current/crystalline axes tilting angle, a constant for all magnetic field. We have plotted $\Delta\psi_m/m$ for (001)- and (111)- surfaces in Fig. 4(a) and (b), respectively. For (001)-surfaces, a phase change for a 4-fold remains a constant of 45°, but those for 2-fold and 6-fold do not. For (111)-surfaces similarly, a phase change for 6-fold stays around the value of 30°, but those for 2-fold and 4-fold do not. As 45° and 30° are the angles that we intentionally tilt the Hall bar patterns for (001)- and (111)-surfaces respectively, it is now apparent that for (001)-surfaces the 4-fold term is solely associated with intrinsic AMR whereas the 2-fold and 6-fold terms are involved with LFE-induced AMR. Similarly, for (111)-surfaces the 6-fold term is solely associated with intrinsic AMR whereas the 2-fold and 4-fold terms are involved with LFE-induced AMR. Therefore, to obtain intrinsic AMR, we need only to look at the non LFE related components. .

We plot LFE-free component $\Delta AMR'_{001}$, defined by $\Delta AMR(\phi)-A_2\exp(-2i\phi)-A_6\exp(-6i\phi)$, as a function of $\phi$ with various fields at 4.2 K for (001) surfaces, for the two current orientations in Fig. 5(b) and (c). The 4-fold symmetric curves with maxima along [110] and [-110] orientations and minima along [001] and [010] orientations are seen in both samples, with a saturation behavior with $H$ at ~ 4 T. As we expected, these curves are bound to crystalline axes and conveys the 4-fold crystalline symmetry for STO (001) surface as illustrated in Fig. 5(a). For (111) surfaces, similarly, we plot the LFE-free $\Delta AMR'_{111}$, defined by $\Delta AMR(\phi)-A_2\exp(-2i\phi)-A_4\exp(-4i\phi)$, as a functions of $\phi$ with various fields at 4.2 K in Fig. 5(e) and (f), with both current directions as shown in the insets. These curves show clearly a roughly 6-fold symmetric and bound to crystalline axes, with maxima pointing to [11-2], [1-21] and [-211] orientations and minima pointing to [-110], [-101] and [0-11] orientations. These generally convey the 6-fold crystalline symmetry for STO (111) surfaces as shown in Fig. 5(d). Interestingly, two peaks at [1-21] are stronger than other four peaks along equivalent directions, in line with Fig. 2(e) and (f), indicating a 6-fold-2-fold symmetry breaking. As the directions of the stronger peaks follows [1-21] axes, and does not depend on current directions, the symmetry breaking cannot be caused by LFE or substrate miscut. The established ordering behind such symmetry breaking is yet to be investigated.

The above results show that the in-plane AMRs for Ar$^+$-irradiated STO surfaces with (001), (110), and (111) orientations all exhibit components that convey the corresponding surface crystalline symmetries, which are 4-fold, 2-fold, and 6-fold respectively. Since our irradiated STO are non-FM as evidenced by the positive MR, the AMR cannot be a consequence of magnetism. On the other hand, Rashba effect causes spin splitting and spin-orbit textures on the 2D Fermi surfaces.[21,24] The symmetries of these 2D Fermi surfaces are 4-fold, 2-fold, and 6-fold for STO (001)-, (110)-, and (111)-surfaces, both obtained by first-principle calculations [21,29,30] and observed by ARPES measurements.[21,29,30,32] According to Boltzmann theory, the conductivity of a metal is determined by the structure of the Fermi surface.[41] Therefore, the symmetries of AMR we observed is most likely a consequence of symmetries of polarized 2D Fermi surfaces. Since Rashba effect only happens on the surfaces/interfaces, the crystalline symmetry associated AMR should disappear when the electron gases lose their 2D nature. Indeed, as we discussed earlier, it is only observed in 2D LAO/STO (001) and is absent in 3D LAO/STO (001) [37]. Also, for Ar$^+$-irradiated STO surfaces with the $n_{2D}$ three orders of magnitude larger than our case, it is not present either.[35] Recently, the Rashba picture for spin splitting was challenged by the observation of a giant spin splitting of ~ 90 meV for 2DEG at STO (001) surfaces, far larger than what is anticipated in the Rashba picture[42]. We note that the 4-fold symmetric AMR for STO (001) surfaces is not likely a result of such giant spin-splitting as the co-centric nature of Ti $3d_{xy}$ band for STO 2DEG does not carry crystalline symmetry information. Rather, it is more likely a result of spin splitting on Ti $3d_{xz}/d_{yz}$ bands that are 4-fold symmetric.

**IV. Conclusion**

We have created metallic states with high mobility on the surfaces of SrTiO$_3$ (STO) with (110), and (111) orientations using Ar$^+$ irradiation and studied their magnetotransport properties for the first time. STO (001) was also studied for comparison. The in-plane anisotropic magnetoresistance (AMR) for these surfaces contains two components: one associated with the crystalline axes and the other from Lorentz force effect. To separate these components, we measured AMR with current along different crystalline axes and performed Fourier analysis. The AMR associated with the crystalline-axes can be separated and we have found symmetries of 4-fold, 2-fold, and nearly 6-fold for (001)-, (110)-, and (111) surfaces respectively. An additional-2-fold symmetry component was observed in the AMR component for (111)-surfaces which breaks the 6-fold symmetry. The symmetries of AMR are most likely caused by Rashba spin-splitting on STO surfaces, a bridging mechanism causing AMR following the symmetries of 2D Fermi surfaces. Our results demonstrated a transport measurement which reflects the symmetry of 2D electronic structures at the STO surfaces.

Acknowledgements

The work was supported in part by the DOE (Grant No. DE-FG02-08ER4653) on measurements and the NSF (Grant No. DMR-1411166) on nanofabrications.

**References**

[1] D. W. Reagor and V.Y. Butko, *Nat. Mater.* **4**, 593 (2005).

[2] A. Ohotomo and H.Y. Hwang, *Nature* **427**, 423 (2004).

[3] A. Tsukazaki, A. Ohtomo, and M. Kawasaki, *Appl. Phys. Lett.* **88**, 152106 (2006).

[4] A. Tsukazaki, A. Ohtomo, T. Kita, Y. Ohno, H. Ohno, and M. Kawasaki, *Science* **315**, 1388 (2007).

[5] A. Tsukazaki, S. Akasaka, K. Nakahara, Y. Ohno, H. Ohno, D. Maryenko, A. Ohtomo and M. Kawasaki, *Nat. Mater.* **9**, 889 (2010).

[6] N. Reyren, S. Thiel, A.D. Caviglia, L.F. Kourkoutis, G. Hammerl, C. Richter, C.W. Schneider, T. Kopp, A.-S. Rüetschi, D. Jaccard, M. Gabay, D.A. Muller, J.-M. Triscone, and J. Mannhart, *Science* **317**, 1196 (2007).

[7] K. Ueno, S. Nakamura, H. Shimotani, A. Ohtomo, N. Kimura, T. Nojima, H. Aoki, Y. Iwasa, and M. Kawasaki, *Nat. Mater.* **7**, 855 (2008).

[8] A.D. Caviglia, S. Gariglio, N. Reyren, D. Jaccard, T. Schneider, M. Gabay, S. Thiel, G. Hammerl, J. Mannhart, and J.-M. Triscone, *Nature* **456**, 624 (2008).

[9] K. Ueno, S. Nakamura, H. Shimotani, H. T. Yuan, N. Kimura, T. Nojima, H. Aoki, Y. Iwasa, and M. Kawasaki, *Nat. Nanotechnol.* **6**, 408 (2011).

[10] A. Brinkman, M. Huijben, M. Van Zalk, J. Huijben, U. Zeitler, J.C. Maan, W.G. Van der Wiel, G. Rijnders, D.H.A. Blank, and H. Hilgenkamp, *Nat. Mater.* **6**, 493 (2007).

[11] L. Li, C. Richter, J. Mannhart, and R.C. Ashoori, *Nat. Phys.* **7**, 762 (2011).

[12] J.A. Bert, B. Kalisky, C. Bell, M. Kim, Y. Hikita, H.Y. Hwang, and K.A. Moler, *Nat. Phys.* **7**, 767 (2011).

[13] Ariando, X. Wang, G. Baskaran, Z.Q. Liu, J. Huijben, J.B. Yi, A. Annadi, A.R. Barman,

A. Rusydi, S. Dhar, Y.P. Feng, J.Ding, H.Hilgenkamp, and T. Venkatesan, *Nat. Commun.* **2**, 188 (2011).

[14] W. Meevasana, P.D.C. King, R.H. He, S-K. Mo, M. Hashimoto, A. Tamai, P. Songsiriritthigul, F. Baumberger, and Z-X. Shen, *Nat. Mater.* **10**, 114 (2011).

[15] J. Biscaras, N. Bergeal, A. Kushwaha, T. Wolf, A. Rastogi, R.C. Budhani, and J. Lesueur, *Nat. Commun.* **1**, 89 (2010).

[16] Y. Chen, N. Pryds, J.E. Kleibeuker, G. Koster, J. Sun, E. Stamate, B. Shen, G. Rijnders, and S. Linderoth, *Nano Lett.* **11**, 3774 (2011).

[17] D. Fuchs, R. Schäfer, A. Sleem, R. Schneider, R. Thelen, and H. von Löhneysen, Appl. Phys. Lett. **105**, 092602 (2014).

[18] Y. Kozuka, M. Kim, C. Bell, B.G. Kim, Y. Hikita, and H.Y. Hwang, *Nature* **462**, 487 (2009).

[19] A. F. Santander-Syro, O. Copie, T. Kondo, F. Fortuna, S. Pailhès, R. Weht, X.G. Qiu, F. Bertran, A. Nicolaou, A. Taleb-Ibrahimi, P. Le Fèvre, G. Herranz, M. Bibes, N. Reyren, Y. Apertet, P. Lecoeur, A. Barthélémy, and M.J. Rozenberg, *Nature* **469**, 189 (2011).

[20] N. Nakagawa, H.Y. Hwang, and D.A. Muller, *Nat. Mater.* **5**, 204 (2006).

[21] P.D.C. King, S.M. Walker, T. Tamai, A. de la Torre, T. Eknapakul, P. Buaphet, S.-K. Mo, W. Meevasana, M.S. Bahramy, and F. Baumberger, *Nat. Commum.* **5**, 3414 (2014).

[22] Z.S. Popovic, and S. Satpathy, *Phys. Rev. Lett.* **94**, 17805 (2005).

[23] K. Janicka, J.P. Velev, and E.Y. Tsymbal, *Phys. Rev. Lett.* **102**, 106803 (2009).

[24] Z. Zhong, A. Toth, and K. Held, *Phys. Rev. B* **87**, 161102(R) (2013).

[25] D. Xiao, W. Zhu, Y. Ran, N. Nagaosa, and S. Okamoto, *Nat. Commun.* **2**, 596 (2011).

[26] A. Rüegg and G.A. Fiete, *Phys. Rev. B* **84** 201103 (2011).

[27] K.-Y. Yang, W. Zhu, D. Xiao, S. Okamoto, Z. Wang, and Y. Ran, *Phys. Rev. B* **84**, 201104 (2011).

[28] D. Doennig, W.E. Pickett, and R. Pentcheva, *Phys. Rev. Lett.* **111**, 126804 (2013).

[29] S.M. Walker, A. de la Torre, F.Y. Bruno, A. Tamai, T.K. Kim, M. Hoesch, M. Shi, M.S. Bahramy, P.D.C. King, and F. Baumberger, *Phys. Rev. Lett.* **113**, 177601 (2014).

[30] T.C. Rödel, C. Bareille, F. Fortuna, C. Baumier, F. Bertran, P. Le Fèvre, M. Gabay, O.H. Cubelos, M.J. Rozenberg, T. Maroutian, P. Lecoeur, and A.F. Santander-Syro, *Phys. Rev. Appl.* **1**, 051002 (2014).

[31] C. Bareille, F. Fortuna, T.C. Rödel, F. Bertran, M. Gabay, O.H. Cubelos, A. Taleb-Ibrahimi, P. Le Fèvre, M. Bibes, A. Barthélémy, T. Maroutian, P. Lecoeur, M.J. Rozenberg, and A.F. Santander-Syro, *Sci. Rep.* **4**, 3586 (2014).

[32] Z. Wang, Z. Zhong, X. Hao, S. Gerhold, B. Stöger, M. Schmid, J. Sánchez-Barriga, A. Varykhalov, C. Franchini, K. Held, and U. Diebold, *PNAS* **111**, 3933 (2014).

[33] G. Herranz, F. Sánchez, N. Dix, M. Scigaj, and J. Fontcuberta, *Sci. Rep.* **2**, 758 (2012).

[34] A. Kalabukhov, R. Gunnarson, J. Börjesson, E. Olsson, T. Claeson, and G. Winkler, *Phys. Rev. B* **75**, 121404(R) (2007).

[35] F.Y. Bruno, J. Tornos, M.G. del Olmo, G.S. Santolino, N.M. Nemes, M. Garcia-Hernandez, B. Mendez, J. Piqueras, G. Antorrena, L. Morellón, J.M. De Teresa, M. Clement, E. Iborra, C. Leon, and J. Santamaria, *Phys. Rev. B* **83**, 245120 (2011).

[36] J.H. Ngai, Y. Segal, F.J. Walker, and C.H. Ahn, *Phys. Rev. B* **83**, 045304 (2011).

[37] A. Annadi, Z. Huang, K. Gopinadhan, X.R. Wang, A. Srivastava, Z.Q. Liu, H.H. Ma,

T.P. Sarkar, T. Venkatesan, and Ariando, Phys. Rev. B **87**, 201102(R) (2013).

[38] M.B. Shalom, C.W. Tai, Y. Lereah, M.Sachs, E. Levy, D. Rakhmilevitch, A. Palevski, and Y. Dagan, *Phys. Rev. B* **80**, 140403(R) (2009).

[39] L. Miao, P. Silwal, X. Zhou, I. Stern, J. Peng, W. Zhang, L. Spinu, Z.Q. Mao, and D.H. Kim, *Appl. Phys. Lett.* **100**, 052401 (2012).

[40] L. Miao, H. Xu, and Z.Q. Mao, *Phys. Rev. B* **89**, 035109 (2014).

[41] K.M. Seemann, F. Freimuth, H. Zhang, S. Blügel, Y. Mokrousov, D.E. Bürgler, and C.M. Schneider, *Phys. Rev. Lett.* **107**, 086603 (2011).

[42] A.F. Santander-Syro, F. Fortuna, C. Bareille, T.C. Rödel, G. Landolt, N.C. Plumb, J.H. Dil, and M. Radovic, *Nat. Mater.* **13**, 1085 (2014).

**Figure Captions**

Figure 1. (Color online) (a) Optical microphotograph of a Hall bar device of irradiated STO (001) surface, with conducting channel highlighted as region A. (b) Sheet resistance $R_{2D}$, (c) 2D carrier density $n_{2D}$, and (d) Hall mobility $\mu_H$ as functions of temperatures for STO (111)-, (110)-, and (001)-surfaces.

Figure 2. (Color online) Relative anisotropic magnetoresistance $\Delta AMR$ as functions of azimuthal angle $\phi$ measured with various magnetic fields $H$ at 4.2 K for STO (001)-surfaces with current along (a) [100], and (b) [110], STO (110)-surfaces with current along (c) [-110], and (d) [001], and STO (111)-surfaces with current along (e) [11-2] and (f) [-101] directions. The insets show the directions of current, $H$, and crystalline axes, as well as the definition of $\phi$.

Figure 3. (Color online) Absolute values of Fourier complex amplitude $A_m$ of relative anisotropic magnetoresistance $\Delta AMR$ as functions of magnetic fields $H$ at 4.2 K for STO (001)-surfaces with current along (a) [100], and (b) [110], STO (110)-surfaces with current along (c) [-110], and (d) [001], and STO (111)-surfaces with current along (e) [11-2] and (f) [-101] directions. The insets show the directions of current, $H$, and crystalline axes.

Figure 4. (Color online) Phase change for the $m$-fold components $\Delta\psi_m$ of relative anisotropic magnetoresistance $\Delta AMR$ caused by the in-plane crystal rotation divided by $m$ as functions of magnetic field *H* at 4.2 K for STO (a) (001)- and (b) (111)-surfaces.

Figure 5. (Color online) Illustration of STO crystalline structure projected onto (a) (001) and (d) (111) surface. Δ*AMR*′$_{001}$ defined by Δ*AMR*(ϕ)-*A*$_2$exp(-2iϕ)-*A*$_6$exp(-6iϕ) for STO (001)-surfaces with current along (b) [100], and (c) [110], and Δ*AMR*′$_{111}$ defined by Δ*AMR*(ϕ)-*A*$_2$exp(-2iϕ)-*A*$_4$exp(-4iϕ) for STO (111)-surfaces with current along (e) [11-2], and (f) [-101] as functions of azimuthal angle ϕ with various magnetic fields *H* at 4.2 K. The insets show the directions of current, *H*, and crystalline axes.

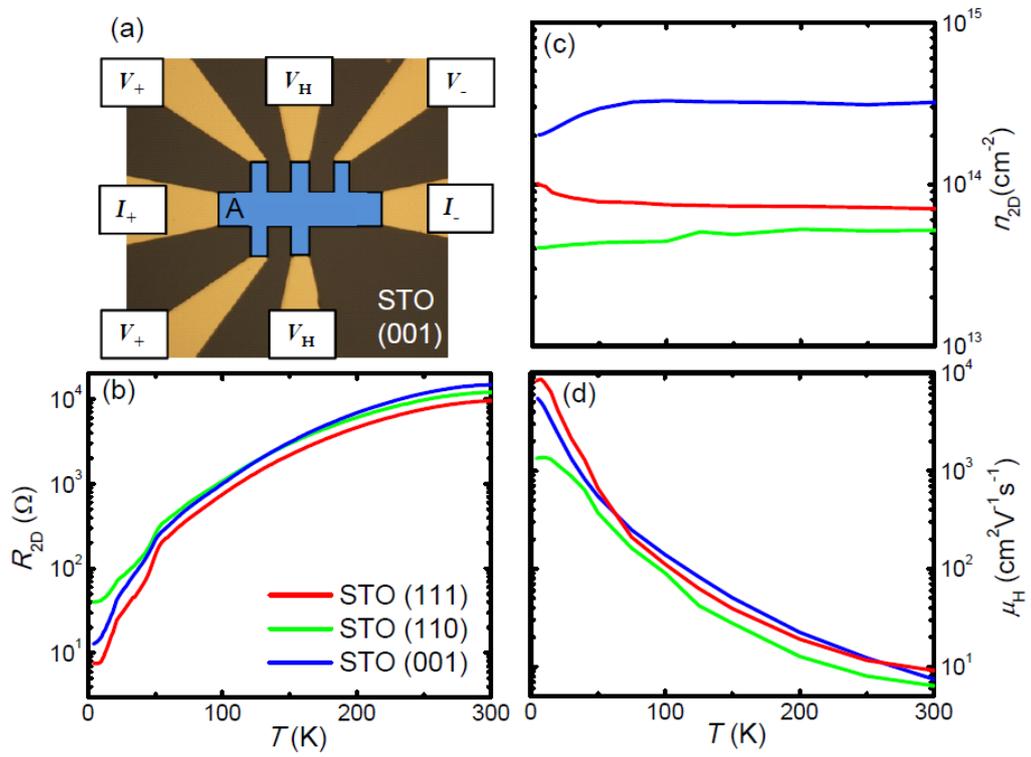

Figure 1, Miao *et al.*

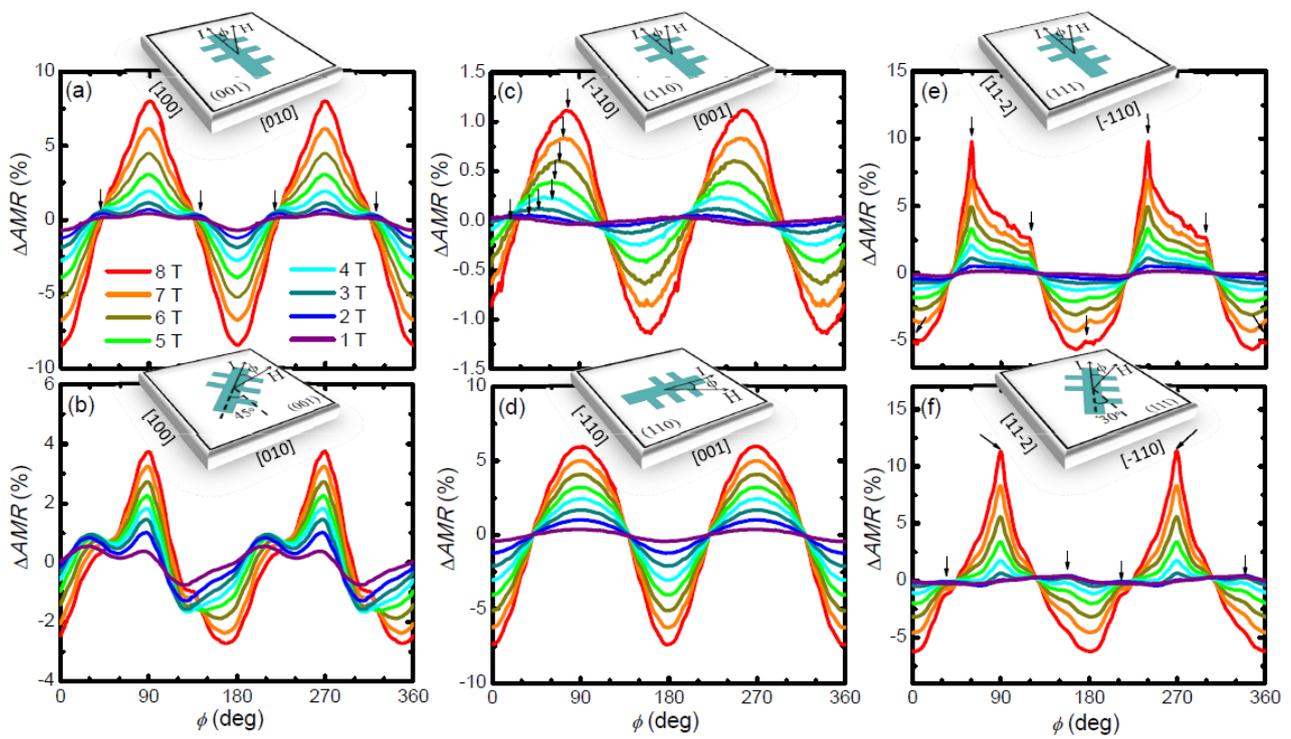

Figure 2, Miao *et al.*

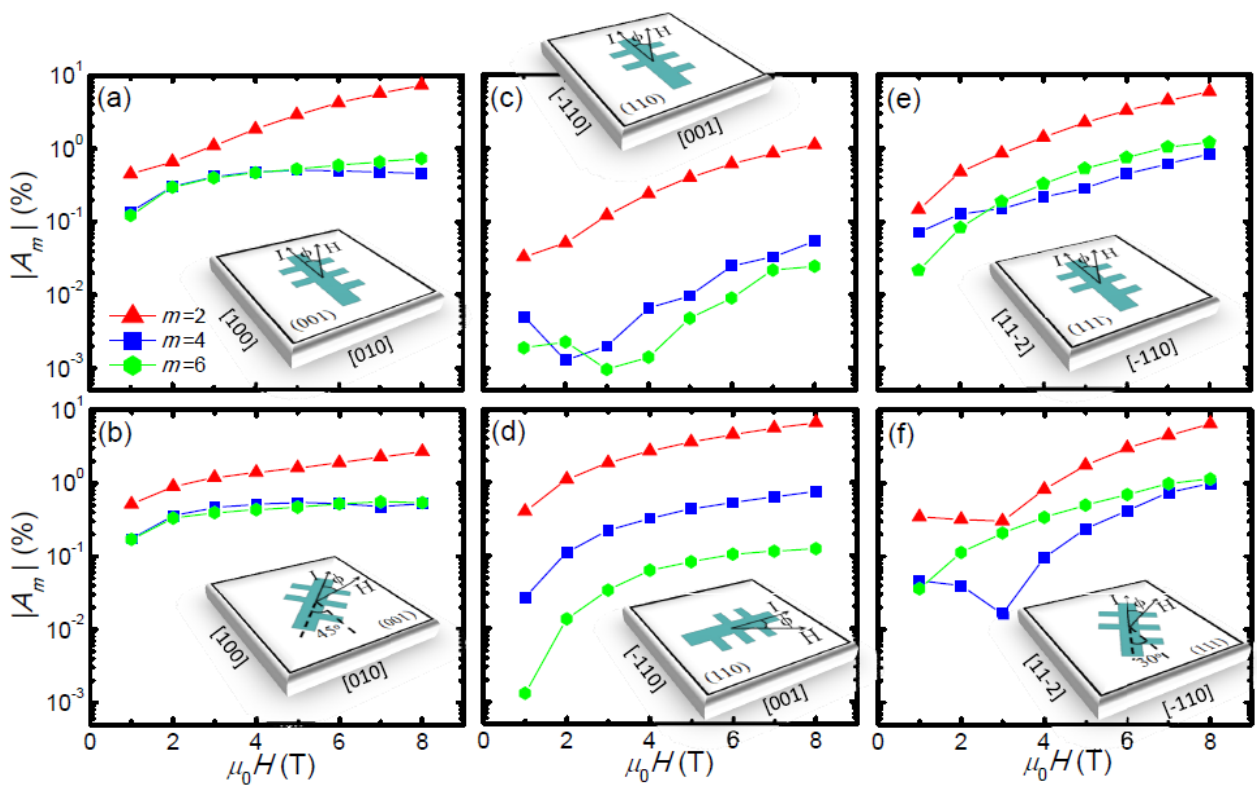

Figure 3, Miao *et al.*

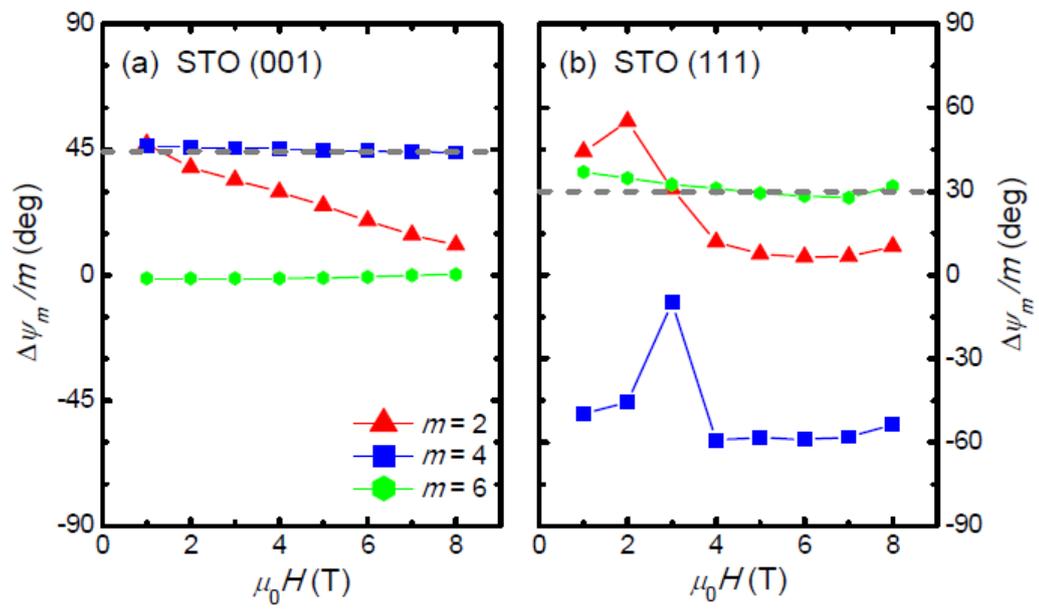

Figure 4, Miao *et al.*

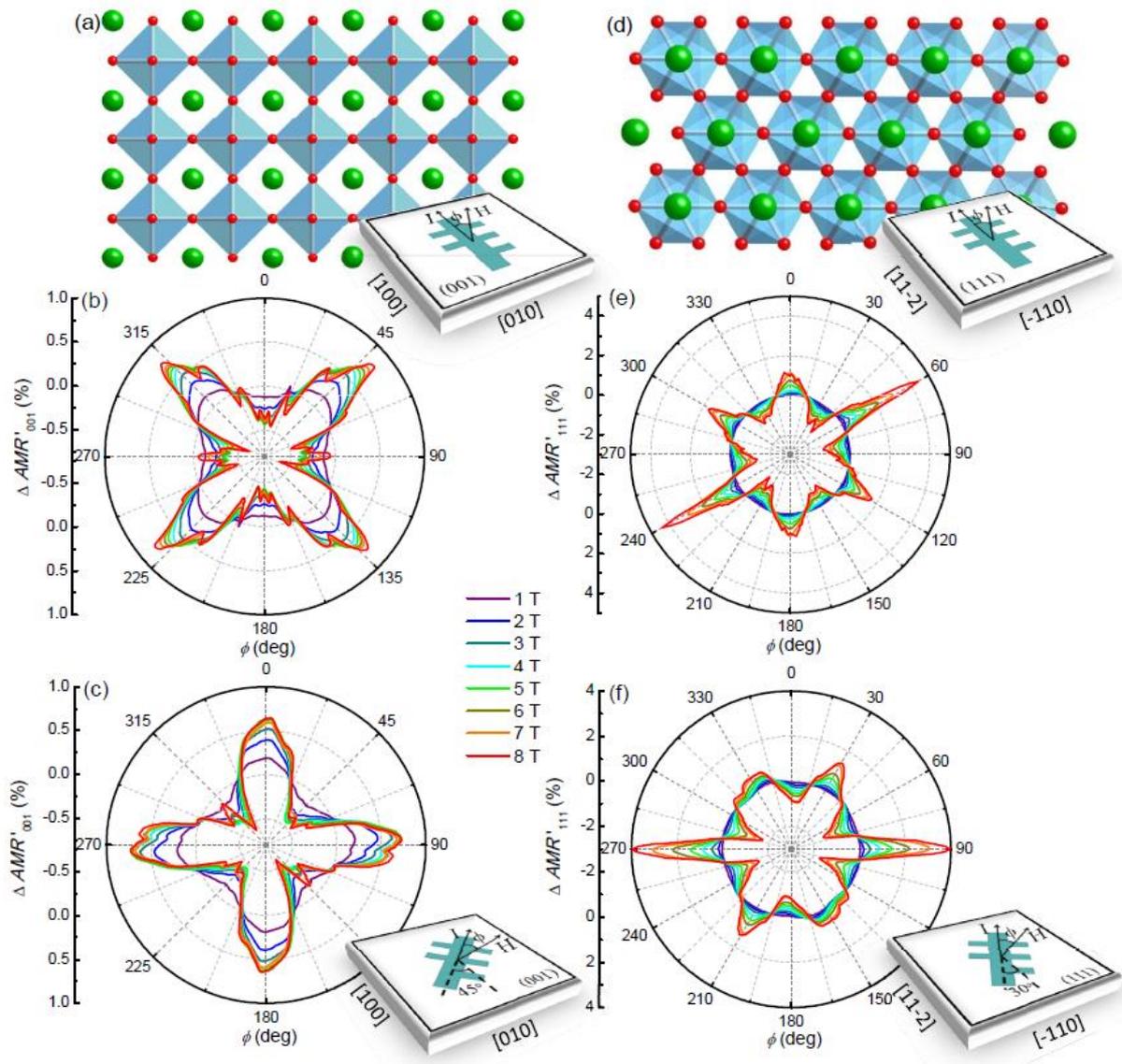

Figure 5, Miao *et al.*